\documentclass{academic}
\usepackage{natbib}
\usepackage{psfig}
\begin{document}

\shortauthor{A.A. Chernova {\em et al.}}
\shorttitle{Response kinetics ...}
\title{Response kinetics of tethered bacteria 
to stepwise changes in nutrient concentration}

\author[1]{Anna A. Chernova}
\author[2]{Judith P. Armitage}
\author[3]{Helen L. Packer} 
\author[4]{Philip K. Maini}

\address[1,4]{Centre for Mathematical Biology,
Mathematical Institute, University of Oxford, 24-29 St Giles', Oxford OX1 3LB, UK}
\address[2]{Microbiology Unit, Department of Biochemistry, University of Oxford, 
South Parks Road, Oxford OX1 3QU, UK}
\address[3]{The School of Biological and Molecular Sciences,
Oxford Brookes University, Headington OX3 0BP, UK}

\correspondence{chernova$@$maths.ox.ac.uk (Anna Chernova)}

\maketitle

\begin{abstract}
We examined the changes in swimming behaviour of the bacterium \textit{Rhodobacter
sphaeroides} in response to stepwise changes in a nutrient (propionate),
following the prestimulus motion, the initial response and the adaptation to the sustained
concentration of the chemical. This was carried out by tethering motile
cells by their flagella to glass slides and following the rotational
behaviour of their cell bodies in response to the nutrient change.
Computerised motion analysis was used to analyse the behaviour.
 Distributions of run and stop times were obtained from rotation data for
 tethered cells. Exponential and Weibull fits for these distributions,
and variability in individual responses are discussed. 

In terms of parameters derived from the run and stop time
distributions, we compare the responses to stepwise changes in the
nutrient concentration and the long-term behaviour of 84 cells under
twelve propionate concentration levels from 1 $nM$ to 25 $mM$.

We discuss traditional assumptions for the random walk approximation to 
bacterial swimming and compare them with the observed 
\textit{R. sphaeroides} motile behaviour.
\end{abstract}


\subsection*{Introduction}

Bacteria populations exhibit intricate spatio-temporal patterns as
a result of reacting actively to environmental cues, such as nutrient
concentration, light levels, oxygen concentration, osmolarity, pH, 
presence of toxins or metabolites, electrochemical stimuli, etc. 
Individual cells integrate these signals to produce a balanced response, 
resulting in active locomotion. 

Bacterial locomotion in general can be achieved by different means,
but most species swim using flagellum-driven motility. 
\textit{Rhodobacter sphaeroides} is a purple non-sulphur 
photosynthetic bacterium. It has a single subpolar flagellum, 
which propels the cell forward, rotating only clockwise.
When the flagellar motor stops, the functional helix relaxes to 
a short wavelength, large amplitude structure and Brownian motion
appears to cause reorientation of the cell body as described in \cite{Arm87}.
When the motor starts to rotate again, a functional helix reforms
and the cell swims in a new direction. Bacterial motion usually
can be described as a random walk with a good approximation. 
From the mathematical point of view it is often assumed that times of 
runs and stops are distributed exponentially.

A wide range of phenomena occur when a signalling cue is 
non-uniformly distributed in space and/or time. 
Such stimuli can lead to directed or biased motion of cells, 
i.e. can cause a tactic response \cite{Schnitzer}. This study is a step towards 
understanding the physical nature of taxis in bacteria 
subjected to changes in chemoattractant concentration. 
We start with a description of the experiment and analysis of 
data and then present results of fitting times of 
runs and stops with a generalisation of the exponential distribution.

\subsection*{Description of the experimental setup}

We are exploiting data for temporal sensing in \textit{R. sphaeroides},
which has been investigated using tethered cells and stepwise changes
in the concentration of proprionate. The data were collected
by H. Packer (Microbiology Unit, Department of Biochemistry,
Oxford University).

\textbf{Growth conditions.} \textit{Rhodobacter sphaeroides} WS8N 
(a wild type nalidixic acid
resistant strain) was grown at \( 30^{o}C \) 
aerobically in the dark as described in \citet{IngArm}.

\textbf{Tethering.} In order to analyse the behaviour of individual bacterial flagellar
motors, cells were tethered by their flagella using antiflagellar
antibody in a flow chamber and the behaviour of individual motors
was analysed using motion analysis. Cells were harvested by gentle centrifugation,
washed and resuspended in nitrogen 10 mM Na HEPES buffer (pH
7.2) containing 50 g/ml of chloramphenicol, aerated by shaking at
\( 30^{0}C \). The cells were starved for 45 mins and tethered in
the flow chamber as described in \cite{Arm96, flow_cell}.

\textbf{Motion analysis.} The tethered cells were viewed via a phase-contrast microscope 
 with an attached video camera. Measurement of cell rotation
was made using the AROT7 software on a Bactracker. 
The software can analyse up to 10 cells per field.
Cells that were rotating without touching others were measured. Data
points were taken at the video frame rate (50 Hz for interlaced images)
and the raw data were downloaded as ASCII files for analysis. 

Tethered cells were subject to stepwise addition of propionate at
the twelve levels with concentrations from 1 \( nM \) to 25 \( mM \)
(i.e. from sub-saturating to saturating concentrations)
 over 3-5 minutes time interval. 

\section*{Data analysis}

\begin{figure}[htb]
\centerline{\psfig{file=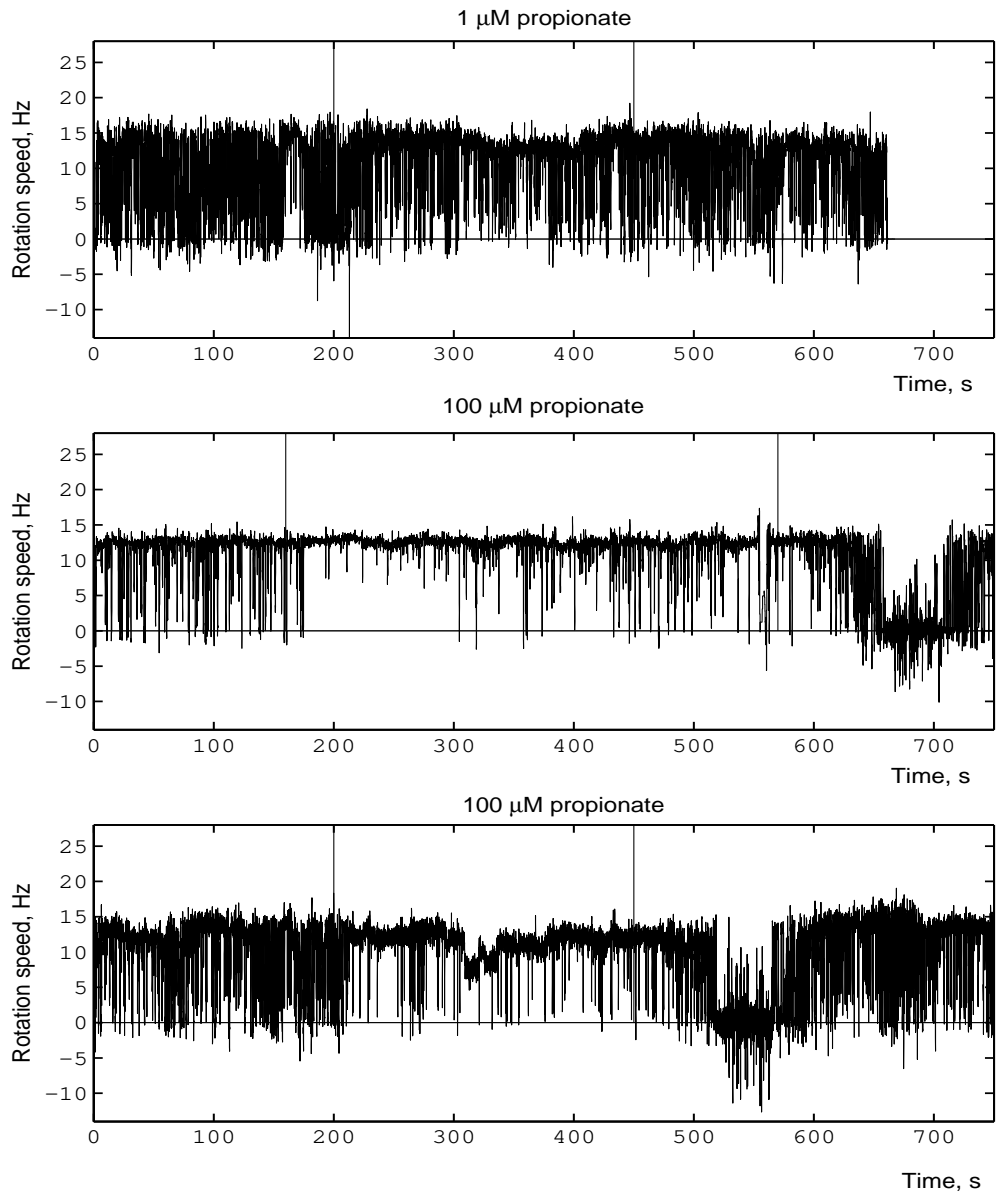,height=16cm,width=13.3cm}}
\caption{Experimental measurements of rotation speed of a single 
tethered \textit{R. sphaeroides} cell. Vertical lines show the time 
intervals during which propionate was applied. Outside these regions 
the propionate concentration is zero. A delay is observed between 
the application of propionate and beginning of response. Duration of 
the delay differs from cell to cell. This part of data was not included 
in the analysis. Data from 1 $\mu M$, 100 $\mu M$ and 1 $mM$ experiments. 
(Unpublished data).}
\label{fig:data}
\end{figure}

During the experiment rotation speed of a single tethered cell was
recorded every 0.02 $s$. Typical time series of responses are shown
in Figure \ref{fig:data}. 
The raw ASCII data files were processed to eliminate system noise, 
each point in the curve was replaced by the weighted average of its
nearest nine neighbours by the method of Savitzky-Golay described in
\cite{SavGol64}.
An advantage of this filter is its sensitivity, so that subtle changes 
in bacterial motor dynamics are preserved in the data. 
The noise arising due to vibrational forces acting on a stopped cell 
must be treated separately.

\begin{figure}[htb]
\centerline{\psfig{file=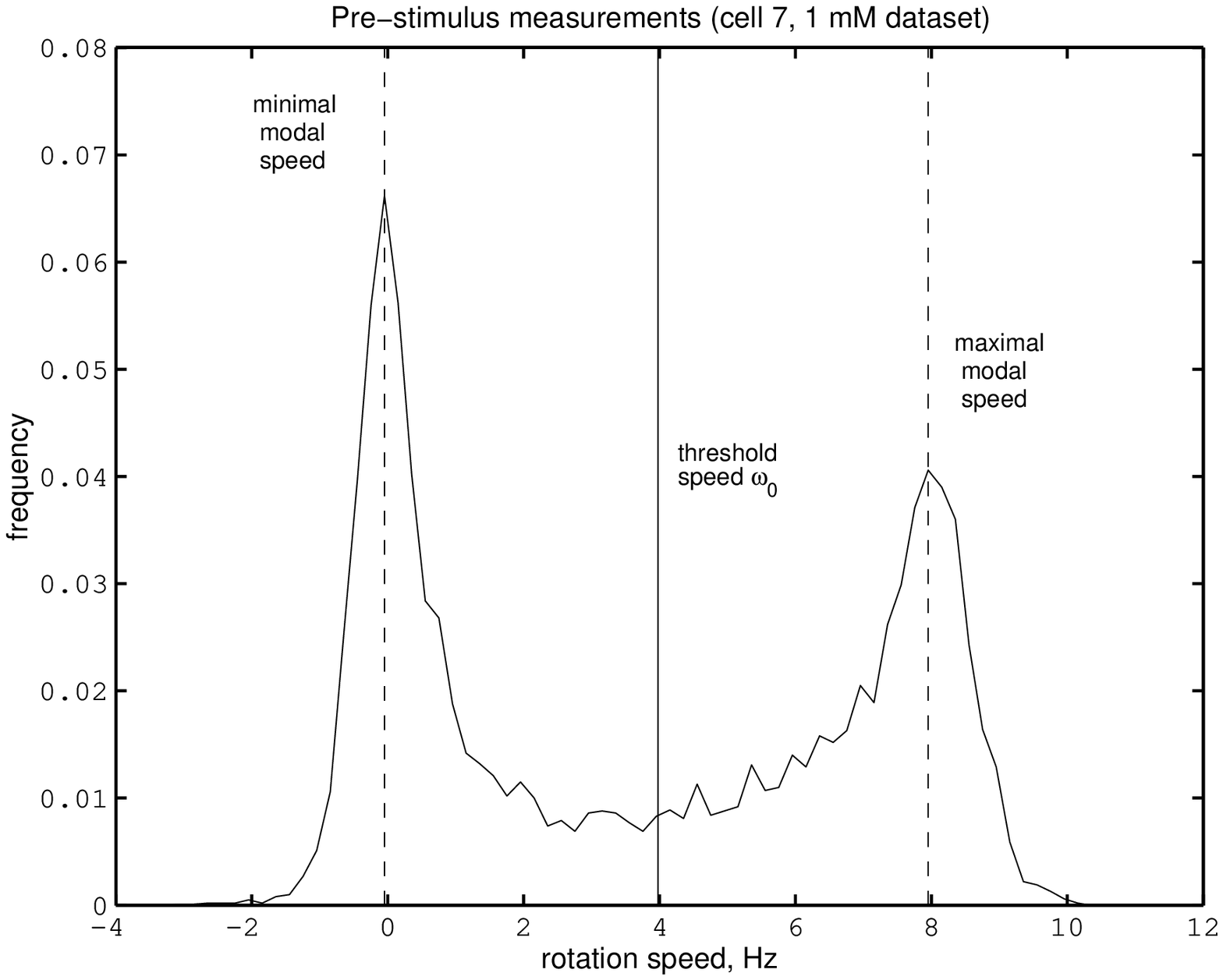,height=8cm,width=10.5cm}}
\caption{ Determination of the threshold level
$\omega_0$ from the distribution of rotation speed data 
(pre-stimulus period). All distributions of the rotation speed 
have two local maxima (two modes). The threshold level was chosen 
to be one half of the largest modal rotation speed.} 
\label{fig:thresh}
\end{figure}

A typical distribution of rotation speed data (pre-stimulus period) 
is shown in Figure \ref{fig:thresh}. The distribution is bimodal.
The first peak around zero corresponds to stops of the motor; the
negative values of the rotation speed appear when the cell, subject to
vibrational forces, turns in the opposite direction.
Vibrational forces contribute in a different manner to the
distribution of speeds around the second maximum. A physical reason
for this is the small value of the Reynold's number for a bacterium in
liquid medium, which means that a moving cell is surrounded by a 
relatively more stable environment than a stopped one.
The complexity of the frequency distribution of the
rotation speed of a moving \textit{R. sphaeroides} cell reflects the
intricate behaviour of the bacterial motor. In comparison with the
virtually constant-speed motor of \textit{Escherichia coli}, in \textit{R.
sphaeroides}, the motor is known to be a variable-speed motor and the
speed variation can be seen both in individual motors and between
cells under the same external conditions. More detail on the
stop$/$start mechanism of the motor in \textit{R. sphaeroides} is given
in \cite{Arm97:rotor}.

For each cell the pre-stimulus records were used to
define a threshold value of rotation speed. From the 
distributions of the rotation speed data we determined the 
two most frequent values (modes); the threshold $\omega_{0}$ 
was chosen to be half of the maximal modal speed 
(Figure \ref{fig:thresh}). 
The cell was considered as \textit{stopped} at time $t$ 
if the detected rotation speed $\omega(t)$ was less 
than $\omega_{0}$, otherwise the cell was 
assumed to be rotating (\textit{running}).

\section*{Distributions of run and stop times}

\begin{figure}[htb]
\centerline{\psfig{file=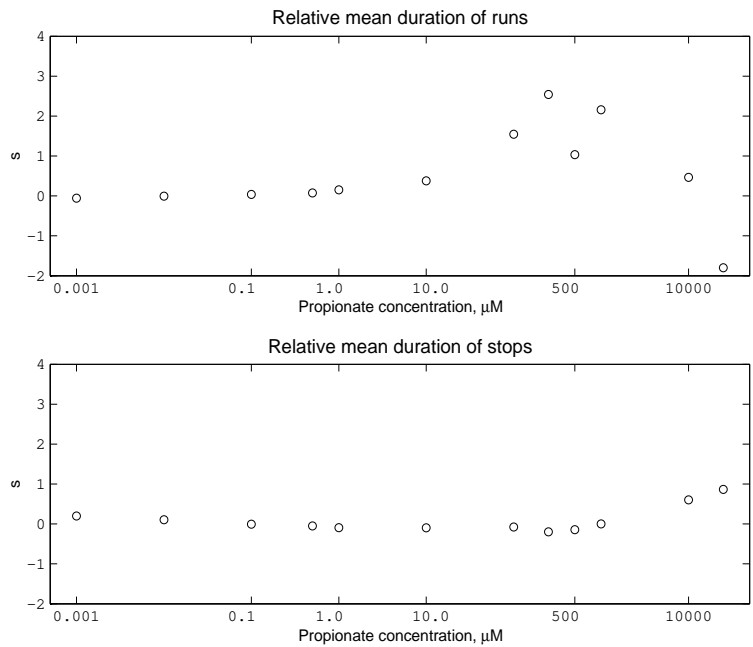,height=8.6cm,width=10cm}}
\caption{Difference between mean values during application
of propionate and pre-stimulus period
\textit{vs.} propionate concentration.}
\label{fig:meantimes_rel}
\end{figure}

Implementing the threshold criteria for the rotation
speed data, we obtain sequences of run and stop times for
each cell.
We computed mean values and standard deviations
of run and stop times averaged over all the cells in one
experiment (i.e. subjected to the same level of propionate)
\textit{vs.} the concentration of propionate. 
Relative mean durations of runs and stops, defined as
differences between mean duration of runs (stops) in the presence
of the nutrient and mean duration of runs (stops) over
pre-stimulus period, are shown in Figure 
\ref{fig:meantimes_rel}.
It was found that on the average the mean duration of
runs remains approximately constant for cells which are subject to
stepwise changes in propionate concentration from zero level to 
some $n_{1}$, $n_{1} < 1$ $\mu M$. 
The stepwise changes from 0 to some value $n_{2}$,
1 $\mu M < n_{2} < 10$ $mM$,
cause an increase in the mean time of runs, while as a 
result of the stepwise changes from 0 to $n_{3}$, 
$n_{3} > 10$ $mM$, the mean duration of runs decreases.
 
The mean duration of stops was not observed to change 
significantly in response to stepwise addition of
propionate at the concentrations from 1 $nM$ to 500 $\mu
M$. When the change was from 0 to some concentration
$n_{4}$, $n_{4} > 500$ $\mu M$, the average duration of stops
increased, but the increase was not larger than 1 $s$.

\begin{figure}[htb]
\centerline{\psfig{file=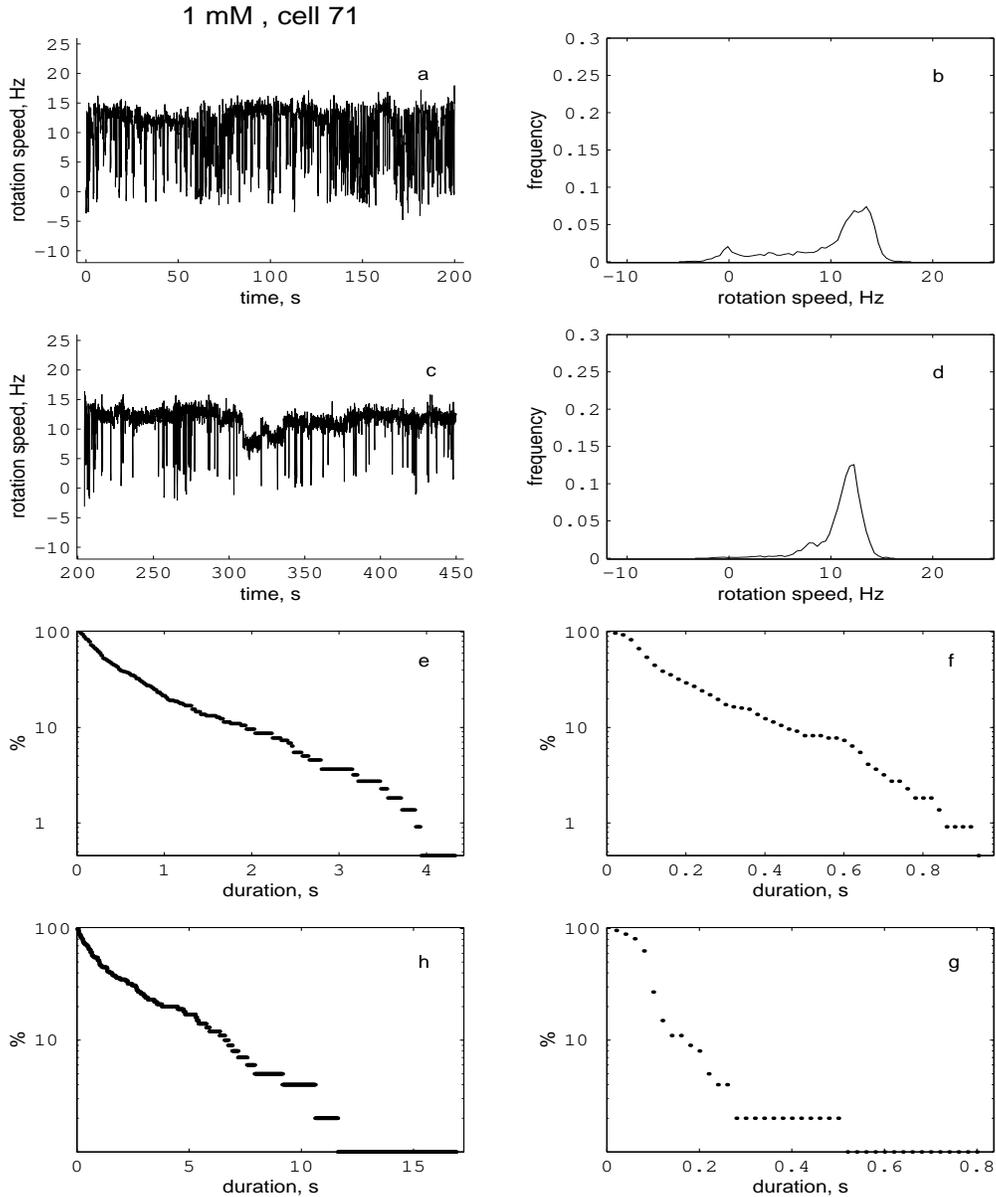,height=16cm,width=13.3cm}}
\caption{{\bf a)} Rotation speed (prestimulus period), 
{\bf b)} frequency distribution of the rotation speed (prestimulus 
period),
{\bf c)} rotation speed (response to $1$ $mM$ propionate), 
{\bf d)} frequency distribution of the rotation speed (response to
$1$ $mM$ propionate). Percentage of {\bf e)} runs (prestimulus
period), {\bf f)} stops (prestimulus period), {\bf h)} runs (response
to 1 $mM$ propionate), {\bf g)} stops (response to 1 $mM$
propionate) as a function of time since the beginning of each 
run{/}stop.}
\label{fig:onecell}
\end{figure}

Figure \ref{fig:onecell} illustrates how a single cell responds to 
stepwise addition and the following new constant level of the nutrient
(here from zero (background) concentration to 1 $mM$):
the cell tends to stop less frequently (the first mode disappears
from the distribution of rotation speeds);
duration of runs increases (up to 4 times in this instance);
there are less long stops than at the background level.

\section*{Fitting with the Weibull distribution function}

In traditional modelling of bacterial motion it is often assumed that 
the temporal part of the random walk can be described by a Poisson process, 
or, more generally, alternating renewal process with exponential 
holding times, i.e. waiting times between runs are distributed exponentially 
(see, for example, \cite{Berg00, Othmer02}).
Here we attempt to justify this assumption by analysing distributions
of both run and stop times obtained from the data.

\begin{figure}[htb]
\centerline{\psfig{file=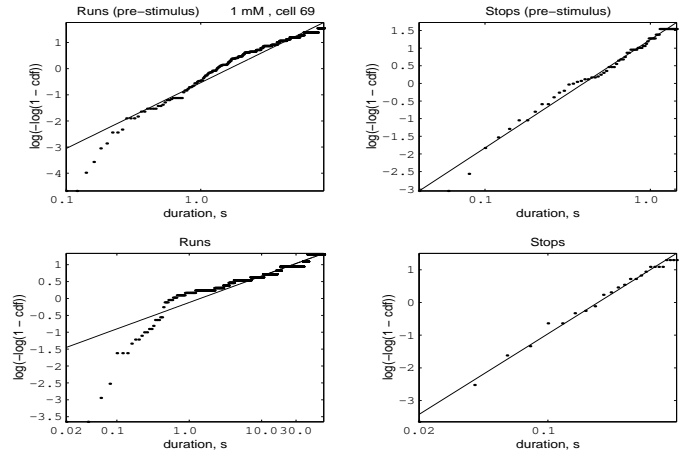,height=6cm,width=9cm}}
\caption{Fitting the Weibull function (\ref{eq:weib})
in the form (\ref{eq:weibline}) to run and stop times of a single
cell during pre-stimulus period (above) and after (below) addition 
of 1 $mM$ propionate.}
\label{fig:weib_one}
\end{figure}

Distributions of run and stop times derived from the data were fitted with
a Weibull distribution function, given by
\begin{equation}
F_w(t)=1 - \exp(-\frac{t^\beta}{{\eta}^{\beta}}),
\label{eq:weib}
\end{equation}
with probability density function
\begin{equation}
f(t)_w=\frac{\beta}{\eta} \Big( \frac{t}{\eta} \Big) ^{\beta-1} \exp(-\frac{t^\beta}{{\eta}^{\beta}}).
\label{eq:weibdensity}
\end{equation}
The Weibull distribution is a generalisation of the exponential
distribution, in the sense that if $\beta=1$ then the former 
reverts to the latter with parameter 
$\mu = \eta$ 
\begin{equation}
\label{eq:exp}
F_e(t)=1 - \exp(-\frac{t}{\mu}).
\end{equation}
A linear equation equivalent to (\ref{eq:weib}) is
\begin{equation}
\label{eq:weibline}
\Phi(\tau)=\beta \tau - \alpha,
\end{equation}
with $\tau=\ln t$, $\Phi=\ln(- \ln(1-F))$ and $\alpha = \beta \ln \eta$. 
An example of fitting data for a single cell is presented in Figure 
\ref{fig:weib_one}. The cumulative distribution function
(cdf) calculated from data was converted into a form suitable
for linear fitting with (\ref{eq:weibline}).

\section*{Results}

\begin{figure}[htb]
\centerline{\psfig{file=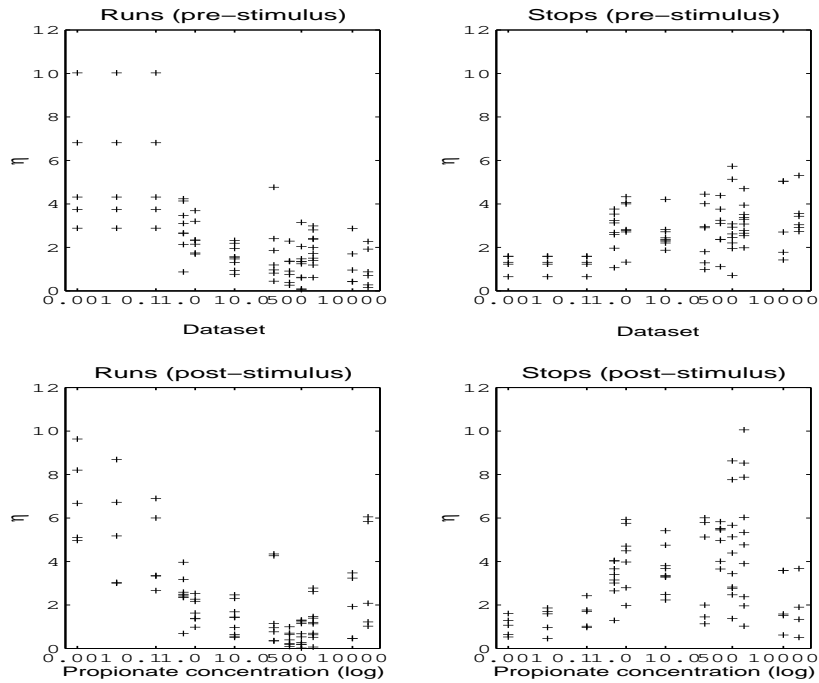,height=9cm,width=11cm}}
\caption{The Weibull parameter $\eta$ estimated for each cell
 in all experiments for run and stop times before and 
 after addition of propionate. The upper plots show the parameters 
 estimated in the absence of nutrient; the values on the horizontal 
 axis correspond to different datasets, according to the subsequent 
 application of propionate. For the three smallest concentration 
 we used the same pre-stimulus segment of the dataset. Two values of $\eta$ are 
not shown in the plot $\eta = 65.04$ estimated for run times
at 25 mM and $\eta = 233.20$ estimated for stop times at 25 mM
(a different cell).}
\label{fig:weib_all_eta}
\end{figure}

\begin{figure}[htb]
\centerline{\psfig{file=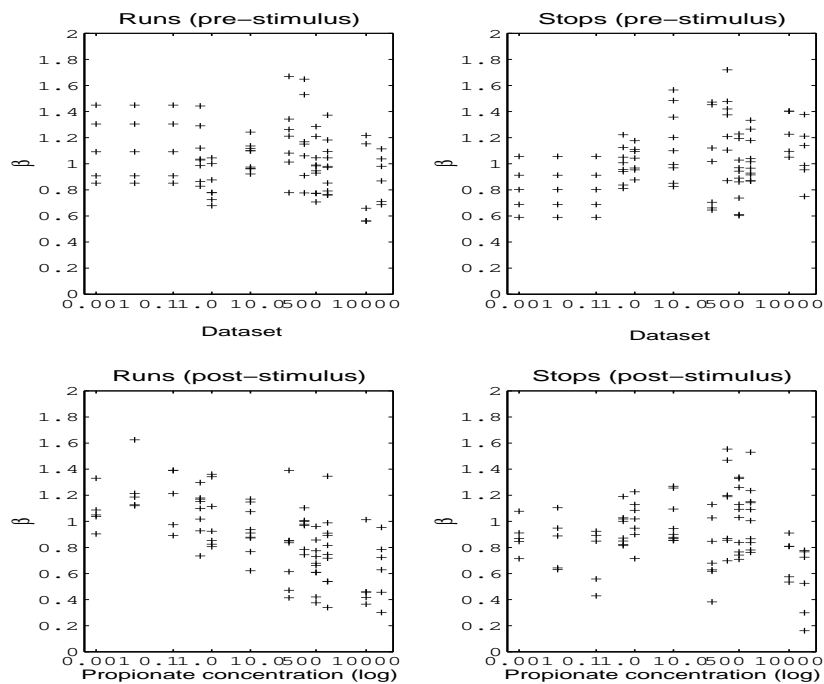,height=9cm,width=11cm}}
\caption{The Weibull parameter $\beta$ estimated for each cell
 in all experiments for run and stop times before and 
 after addition of propionate. When $\beta = 1$,
 the distribution is exponential with coefficient
 equal to the corresponding value of $\eta$. The upper plots show the parameters 
 estimated in the absence of nutrient; the values on the horizontal 
 axis correspond to different datasets, according to the subsequent 
 application of propionate. For the three smallest concentration 
 we used the same pre-stimulus segment of the dataset.}
\label{fig:weib_all_beta}
\end{figure}

Figures \ref{fig:weib_all_eta} and \ref{fig:weib_all_beta} show the parameters
$\eta$ and $\beta$ estimated for each cell in all the experiments.

As can be seen from the Weibull parameter estimation, there is 
a significant variability in the cell behaviour. 97.7 percent of 
all $\eta$ values lie within the range from 0 to 12, and two 
values were found to be 65.04 (for run times, 25 $mM$
propionate) and 223.20 (stop times, 25 $mM$ propionate, different
cell).

The second parameter, $\beta$, significantly deviates from unity 
for both run and stop times before and after addition of nutrient.

One of the possible conclusions from analysis of norm of residuals 
of the fits is that the Weibull distribution is a better description 
for stop times (both in the pre-stimulus period and after change in 
the chemical concentration) than for run times.

The results of the analysis of run and stop times do not allow us to
draw a single conclusion from the data, mainly due to a great 
diversity in individual cells' behaviour. 
Cases of successful fitting of a run$/$stop distribution with Weibull
functions are found as often as cases with poor results of such fitting.
In some instances, like for the distribution of run times for a cell
subjected to 1 $mM$ propionate (Figure \ref{fig:weib_one}), a linear 
combination of Weibull functions could be chosen for fitting.

Preliminary analysis of independence of successive time intervals 
also show a high variability among individual cells, and does not 
allow us to talk about a ``typical behaviour''. On the average, however, 
the autocorrelation function calculated for the sequences of run and 
stop times does not differ significantly from the autocorrelation 
function of a Poisson process.

These complexities discovered in the data suggest that a detailed 
investigation into the run and stop times distributions can produce 
some interesting results. We have chosen the Weibull distribution 
function as a simplest generalisation of the exponential distribution. 
Such generalisation leads, in fact, to re-consideration of several 
modelling hypotheses for chemotaxis in bacteria. Experimental motivation 
of the generalisation and modification of the temporal and spatio-temporal 
models is a subject of our current research.

\section*{Acknowledgements}
The authors thank Dr T. Alarc\'{o}n, Dr R. Satnoianu and Dr D. Sumpter for helpful 
and stimulating discussions.

AAC gratefully acknowledges scholarships from ORS Award Scheme and 
Oxford University Bursary. This worked arose as a part of the project 
supported by the joint BBSRC/EPSRC Initiative in ``Mathematical Modelling, 
Simulation and Prediction of Biological Systems'', grant 43-MMI 09782 awarded 
to J.P. Armitage and P.K.Maini. HLP thanks the NERC for their support.

\end{document}